\documentclass[prl,twocolumn,showpacs,amsmath,amssymb,superscriptaddress]{revtex4}
\usepackage{amssymb}
\usepackage{txfonts}
\usepackage{bbm}
\usepackage{graphicx}
\usepackage{appendix}
\usepackage{epsf}
\usepackage{epstopdf}
\usepackage{amsmath}
\usepackage[usenames]{color}

\begin{document}
\title{Proposed Detection of Time Reversal Symmetry in Topological Surface States}

\author{Degang Zhang}
\affiliation{College of Physics and Electronic Engineering, Sichuan Normal University,
Chengdu 610101, China}
\affiliation{Institute of Solid State Physics, Sichuan Normal
University, Chengdu 610101, China}
\affiliation{Texas Center for Superconductivity and Department of
Physics, University of Houston, Houston, TX 77204, USA}

\author{C. S. Ting}
\affiliation{Texas Center for Superconductivity and Department of
Physics, University of Houston, Houston, TX 77204, USA}

%\date{\today}
\begin{abstract}

By employing T-matrix approach, we investigate a nonmagnetic impurity
located on the surface of three-dimensional topological insulators,
where the time reversal symmetry is preserved. It is shown that
the images of the local density of states (LDOS) around the single
impurity have the dip-hump structures with six-fold symmetry at
different bias voltages. The peaks are produced by quasiparticle interference
while the local minima at the backscattering wave vectors are
due to the absence of backscatterings. With increasing the bias voltage, the
peaks and dips move forward to the location of the impurity.
These dips at the backscattering wave vectors in the LDOS spectra
can be regarded as a signature of
the time reversal symmetry in the topological surface states,
which could be observed by scanning tunneling microscopy.

\end{abstract}


\maketitle

In recent years much attention has been focused on the topological surface states
due to their potential applications in quantum computing or spintronics [1,2].
The Dirac-cone-like electronic states, existing on the surface
of three-dimensional bulk insulating materials, have been observed
in Bi$_{1-x}$Sb$_{x}$ [3], Bi$_2$Sb$_3$ [4], Sb$_2$Te$_3$ [5],
Bi$_2$Te$_3$ [5,6], TlBiSe$_2$ and TlBiTe$_2$ [7], by angle-resolved
photoemission spectroscopy (ARPES).
The surface energy band structure was determined by employing $k\cdot p$ theory [8],
where an unconventional hexagonal warping term plays a crucial role in
fitting the ARPES data [3-7] and in explaining the scanning tunneling microscopy
(STM) experiments [9-13].

It is known that the novel topological surface states possess
time reversal symmetry (TRS), which leads to the absence of backscatterings.
Such a symmetry induces unusual electron transport properties
on the surface of topological insulators [9-13]. Therefore,
it is interesting to explore the TRS in order to understand thoroughly
the physical properties of topological insulators. In this work,
we give a proposal for experimental detection of the TRS in the
topological surface states. We note that the oscillations of
local density of states (LDOS) induced by a step or line defect
can be explained satisfically by quasiparticle interference (QPI) and
seem not to exhibit the existence of the
TRS due to no direct evidence showing the absence of backscatterings. [9-13].
Here we employ a nonmagnetic impurity located on the surface of topological
insulators in order to detect this symmetry. Such a single impurity has
also been used to judge the superconducting order parameter symmetry in
the cuprates and the FeAs-based superconductors [14-16].
In Refs. [17,18], the Fourier transformation of the LDOS
induced by a nonmagnetic impurity was calculated. However,
the oscillations of the LDOS with the hexagonal warping term
and the role of the TRS in real space are never discussed previously.

The momentum space Hamiltonian describing the surface states of
three-dimensional topological insulators reads [8]

$$H_0=\sum_{\bf k}C^\dagger_{\bf k}[(\frac{k^2}{2m^*}-\mu)I+v(k_x\sigma_y-k_y\sigma_x)
 +\lambda \phi_{\bf k}\sigma_z]C_{\bf k},\eqno{(1)}$$
where $C^\dagger_{\bf k}=(c^\dagger_{{\bf k}\uparrow},c^\dagger_{{\bf k}\downarrow})$,
$I$ and $\sigma_i (i=x,y,z)$ are the $2\times 2$ unit matrix
and the Pauli matrices, respectively, $m^*$ is the effective mass of
electrons, which is usually very large for the topological
insulators, $\mu$ is the chemical potential to be determined by doping [10],
$v$ is the strength of the Rashba spin-orbit coupling, the last term is the so called
hexagonal warping term, and $\phi_{\bf k}\equiv k_x(k_x^2-3k_y^2)$.
Taking the linear transformations $c_{{\bf k}\uparrow}=\frac{1}{\sqrt{2}}\sum_{s=0,1}
\alpha_{{\bf k}s}a_{{\bf k}s}\psi_{{\bf k}s}$ and
$c_{{\bf k}\downarrow}=\frac{1}{\sqrt{2}}\sum_{s=0,1}
\beta_{{\bf k}s}a_{{\bf k}s+1}\psi_{{\bf k}s}$
with $\alpha_{{\bf k}s}=1-s+s (ik_x+k_y)k^{-1}$,
$\beta_{{\bf k}s}=s+(1-s) (ik_x-k_y)k^{-1}$,
$a_{{\bf k}s}=cos\frac{\theta_{\bf k}}{2}+(-1)^{s}sin\frac{\theta_{\bf k}}{2}$,
$\theta_{\bf k}=arctan\frac{\lambda\phi_{\bf k}}{vk}$, and $k=\sqrt{k^2_x+k^2_y}$,
we have $H_0=\sum_{{\bf k}s}E_{{\bf k}s}\psi^{\dagger}_{{\bf k}s}\psi_{{\bf k}s}$
with the eigenenergies
$E_{{\bf k}s}=\frac{k^2}{2m^*}+(-1)^s\sqrt{\lambda^2\phi^2_{\bf k}+v^2k^2}-\mu$.

Now we investigate a singe impurity located at the origin ${\bf r}=(0,0)$
on the surface of three-dimensional topological insulators.
Without loss of generality, such an impurity can be described by the Hamiltonian
$H_{\rm imp}=U\sum_{\sigma}c^\dagger_{{\bf 0}\sigma}c_{{\bf 0}\sigma}
+V_{\rm m}(c^\dagger_{{\bf 0}\uparrow}c_{{\bf 0}\uparrow}
-c^\dagger_{{\bf 0}\downarrow}c_{{\bf 0}\downarrow})$. Here $U$ and $V_{\rm m}$
are the nonmagnetic and magnetic parts of the impurity strength, respectively.
We solve the total Hamiltonian $H=H_0+H_{\rm imp}$ by
employing T-matrix approach [15,16] and finally obtain the analytical expression
of the LDOS around the impurity
$$\rho({\bf r},\omega)\equiv \sum_{{\bf k}{\bf k}^\prime ss^\prime}
\rho^{{\bf k}^\prime s^\prime}_{{\bf k}s}({\bf r},\omega)$$
$$=-\frac{1}{2N\pi}{\rm Im}
\sum_{{\bf k}{\bf k}^\prime ss^\prime}e^{i({\bf k}-{\bf k}^\prime)\cdot {\bf r}}
(\alpha_{{\bf k}s}\alpha^*_{{\bf k}^\prime s^\prime}a_{{\bf k}s}a_{{\bf k}^\prime s^\prime}$$
$$+\beta_{{\bf k}s}\beta^*_{{\bf k}^\prime s^\prime}a_{{\bf k}s+1}a_{{\bf k}^\prime s^\prime+1})
G^{{\bf k}^\prime s^\prime}_{{\bf k}s}(i\omega_n)|_{i\omega_n\rightarrow
\omega+i0^+}, \eqno{(2)}$$
where $N$ is the number of lattices, $\rho^{{\bf k}^\prime s^\prime}_{{\bf k}s}({\bf r},\omega)$
represents the LDOS contributed by the scattering process
$({\bf k}s)\rightarrow ({\bf k}^\prime s^\prime)$, and the Green's functions
$G^{{\bf k}^\prime s^\prime}_{{\bf k}s}(i\omega_n)=
G^{0}_{{\bf k}s}(i\omega_n)\delta_{{\bf k}{\bf k}^\prime}\delta_{ss^\prime}+
G^{0}_{{\bf k}s}(i\omega_n){\cal T}^{{\bf k}^\prime s^\prime}_{{\bf k}s}(i\omega_n)
G^{0}_{{\bf k}^\prime s^\prime}(i\omega_n)$ with the bare Green's functions
$G^{0}_{{\bf k}s}(i\omega_n)=(i\omega_n-E_{{\bf k}s})^{-1}$ and the T matrix

$${\cal T}^{{\bf k}^\prime s^\prime}_{{\bf k}s}(i\omega_n)=\frac{U+V_m}{2N}
\frac{\alpha^*_{{\bf k}s}a_{{\bf k}s}
\alpha_{{\bf k}^\prime s^\prime}a_{{\bf k}^\prime s^\prime}}
{[1-(U+V_m)A(i\omega_n)]}$$

$$+\frac{U-V_m}{2N}
\frac{\beta^*_{{\bf k}s}a_{{\bf k}s+1}
\beta_{{\bf k}^\prime s^\prime}a_{{\bf k}^\prime s^\prime+1}}
{[1-(U-V_m)A(i\omega_n)]}.
\eqno{(3)}$$
Here, $A(i\omega_n)=\frac{1}{2N}\sum_{{\bf k}s}G^{0}_{{\bf k}s}(i\omega_n)$.
When $V_m=0$, we have ${\cal T}^{-{\bf k}s}_{{\bf k}s}(i\omega_n)\equiv 0$.
This means that there are no backscatterings at each constant-energy contour
$\omega=E_{{\bf k}s}$ [17,18].
In other words, the surface states with opposite momentum and the same $s$
are still orthogonal and do not mix after being scattered by
a pure nonmagnetic impurity. So there is no interference
between the two surface states connected by a backscattering wave vector.
From Eq. (2), we can also see that the backscattering processes have no contribution
to the LDOS, i.e. $\rho^{-{\bf k}s}_{{\bf k}s}({\bf r},\omega)\equiv 0$.
Therefore, the LDOS induced by a nonmagnetic impurity
should has the local minima at the backscattering wave vectors.
Such the dips are produced by the absence of backscatterings rather than
the QPI, which could be used to detect the TRS in the topological
surface states.

\begin{widetext}

\begin{figure}

\rotatebox[origin=c]{0}{\includegraphics[angle=0,
           height=1.8in]{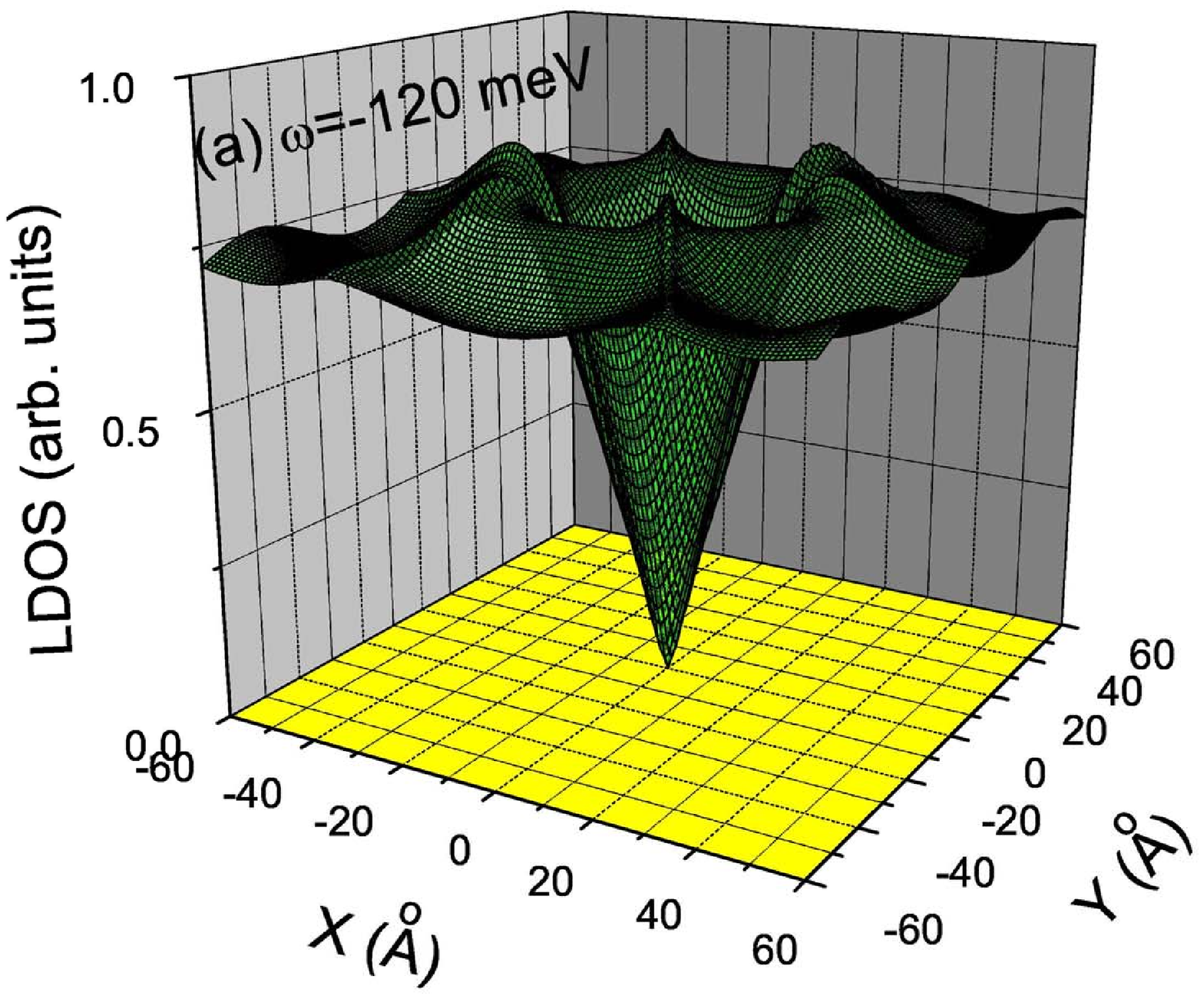}}
\rotatebox[origin=c]{0}{\includegraphics[angle=0,
           height=1.8in]{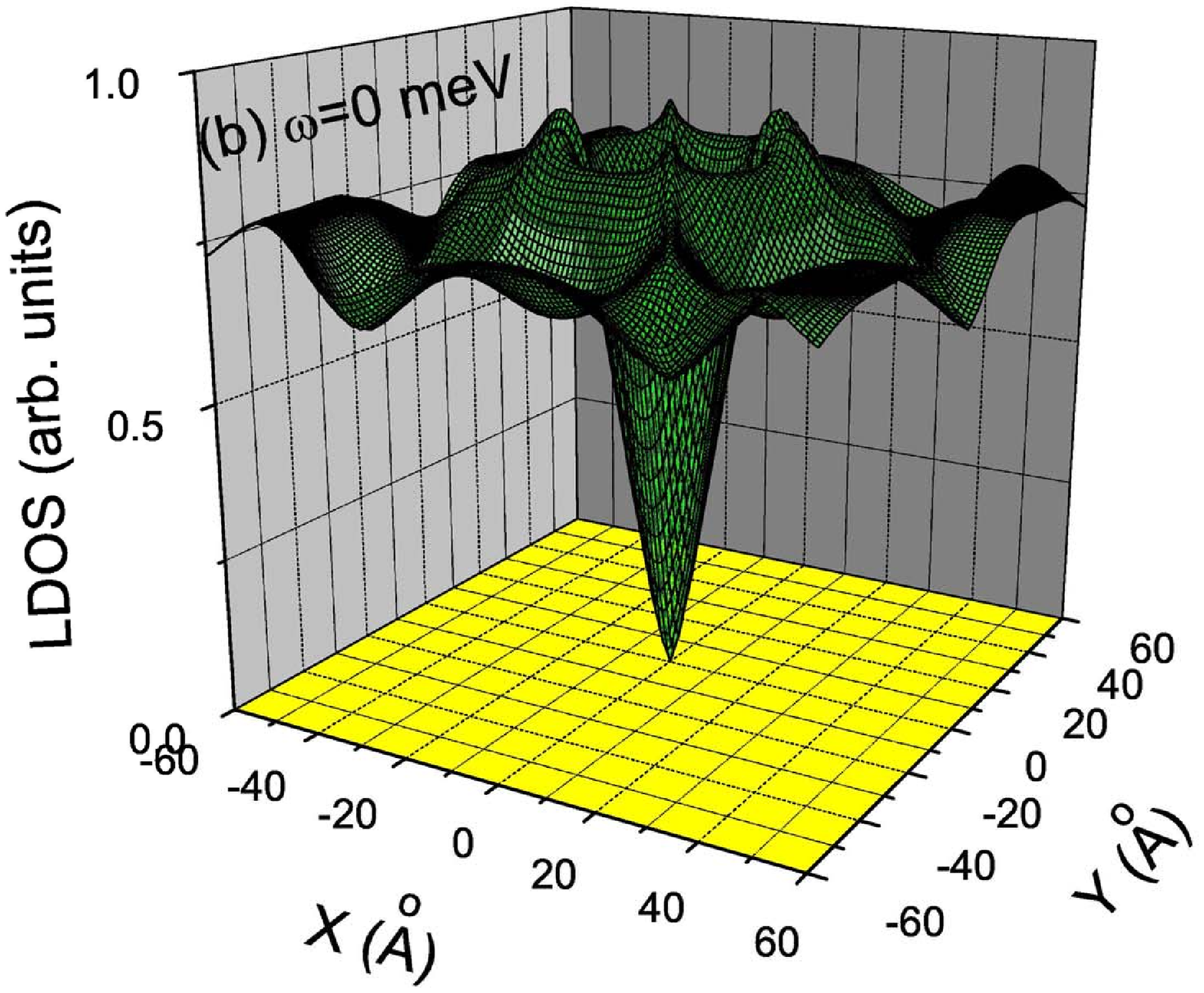}}
\rotatebox[origin=c]{0}{\includegraphics[angle=0,
           height=1.8in]{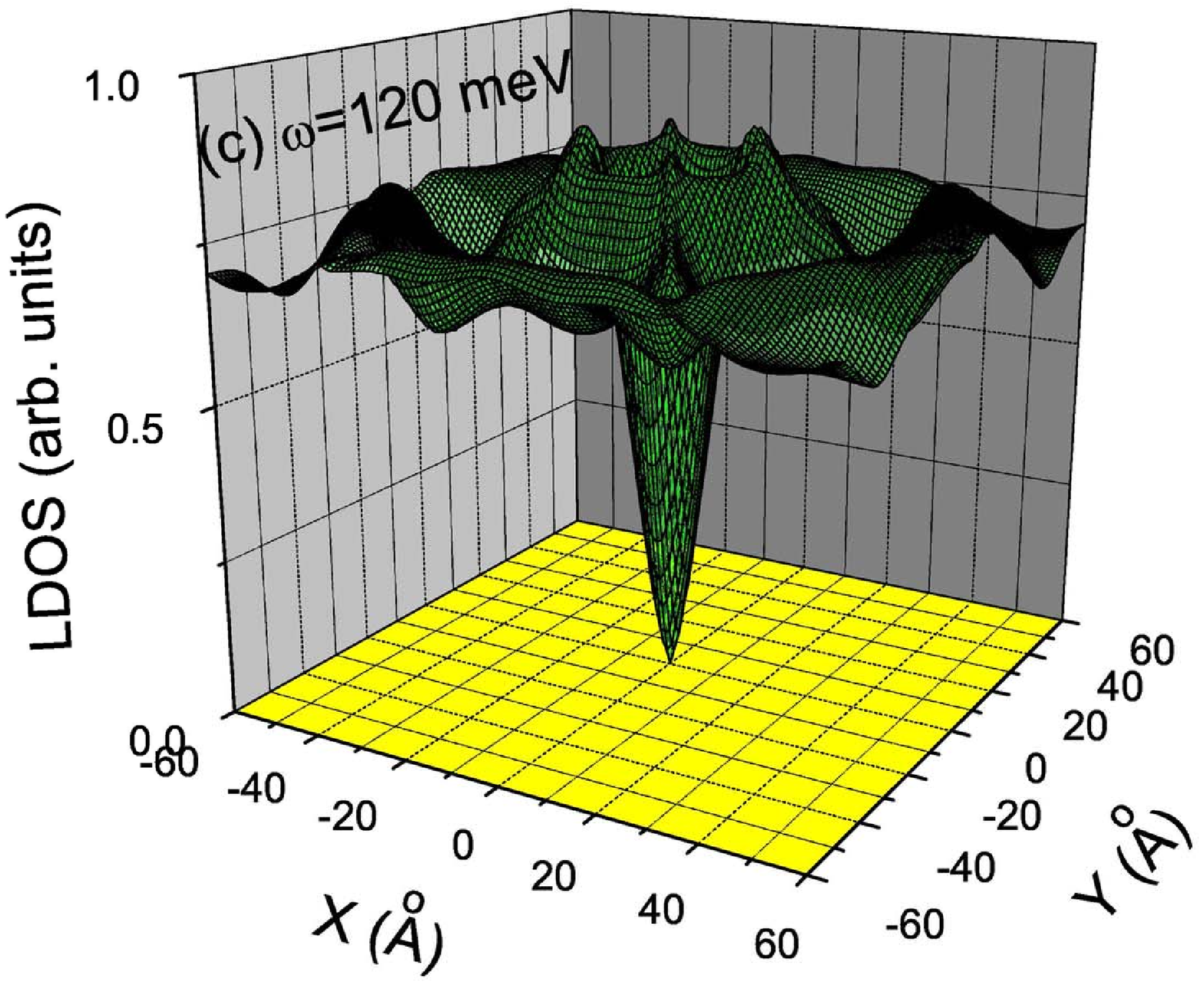}}
\caption{(Color online) The stereographs of the LDOS $\rho({\bf r},\omega)$
around a unitary impurity, i.e. $U\rightarrow \infty$, at different  bias voltages.}
\end{figure}

\end{widetext}

According to Eq. (2), we can calculate the LDOS around the single impurity, which is
measured by STM experiments. In our calculations, we choose a $120\times 120\AA^2$
lattice with $N=1800\times 1800$ and employ the physical parameters
in Bi$_2$Sb$_3$, i.e.  $\lambda=250.0$
eV$\cdot$\AA$^3$, $v=2.55$ eV$\cdot$\AA, and $\mu=0.334$ eV [8,10].

In Fig. 1, we present the stereographs of the LDOS induced by a unitary impurity
($U\rightarrow \infty$) at $\omega=-120, 0, 120$ (meV).
Such the infinite potential has been realized by the Zn ions in the cuprates [14],
which is also expected to exist on the surface of topological insulators.
Obviously, $\rho({\bf r},\omega)$
has a dip-hump structure with a six-fold symmetry at each bias voltage.
When the bias voltage $\omega$ increases, these peaks and dips move forward
to the location of the impurity. On the impurity site, we have
$\rho({\bf r},\omega)|_{{\bf r}=(0,0)}=0$.

\begin{figure}

\rotatebox[origin=c]{0}{\includegraphics[angle=0,
           height=2.2in]{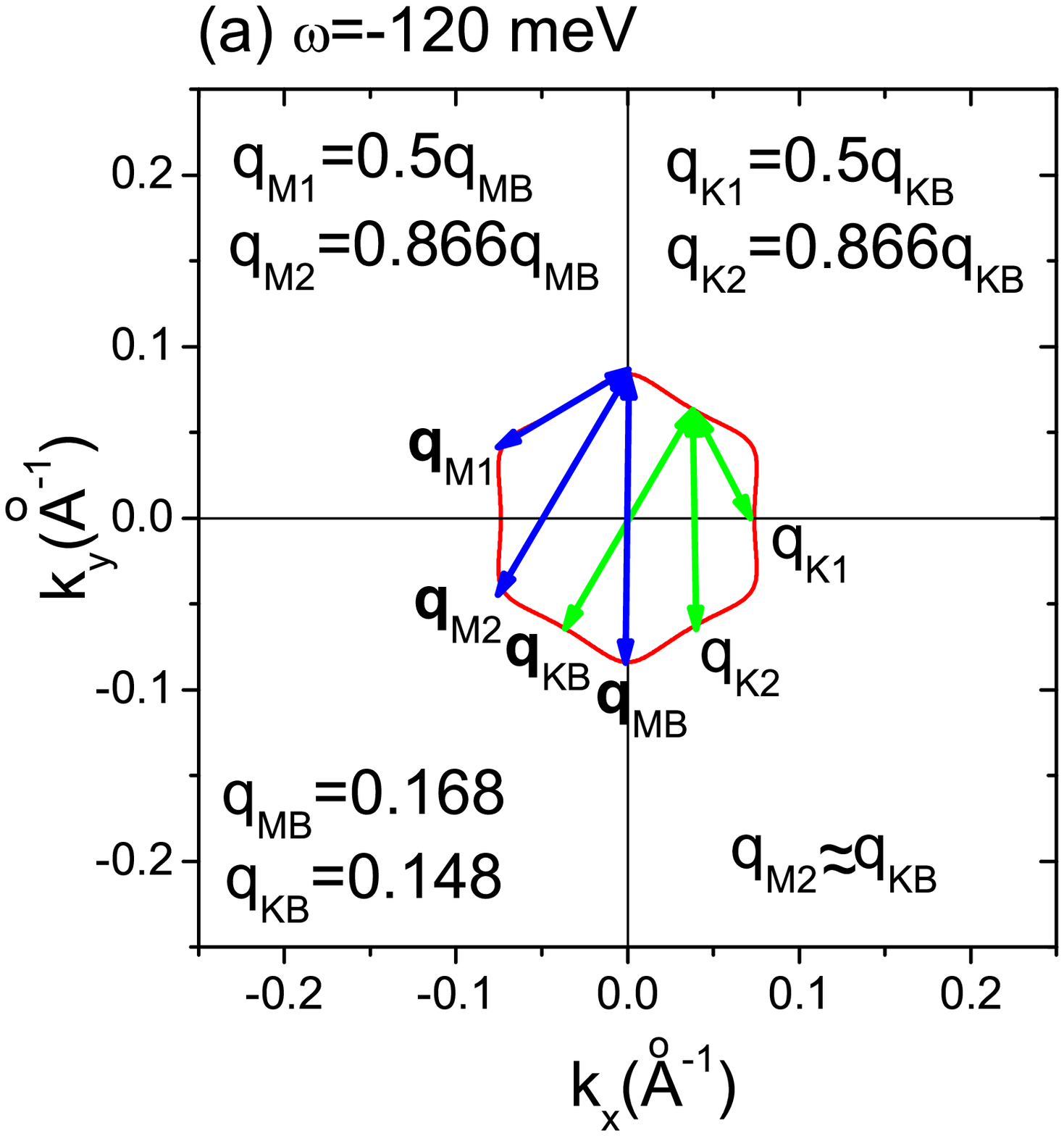}}
\rotatebox[origin=c]{0}{\includegraphics[angle=0,
           height=2.2in]{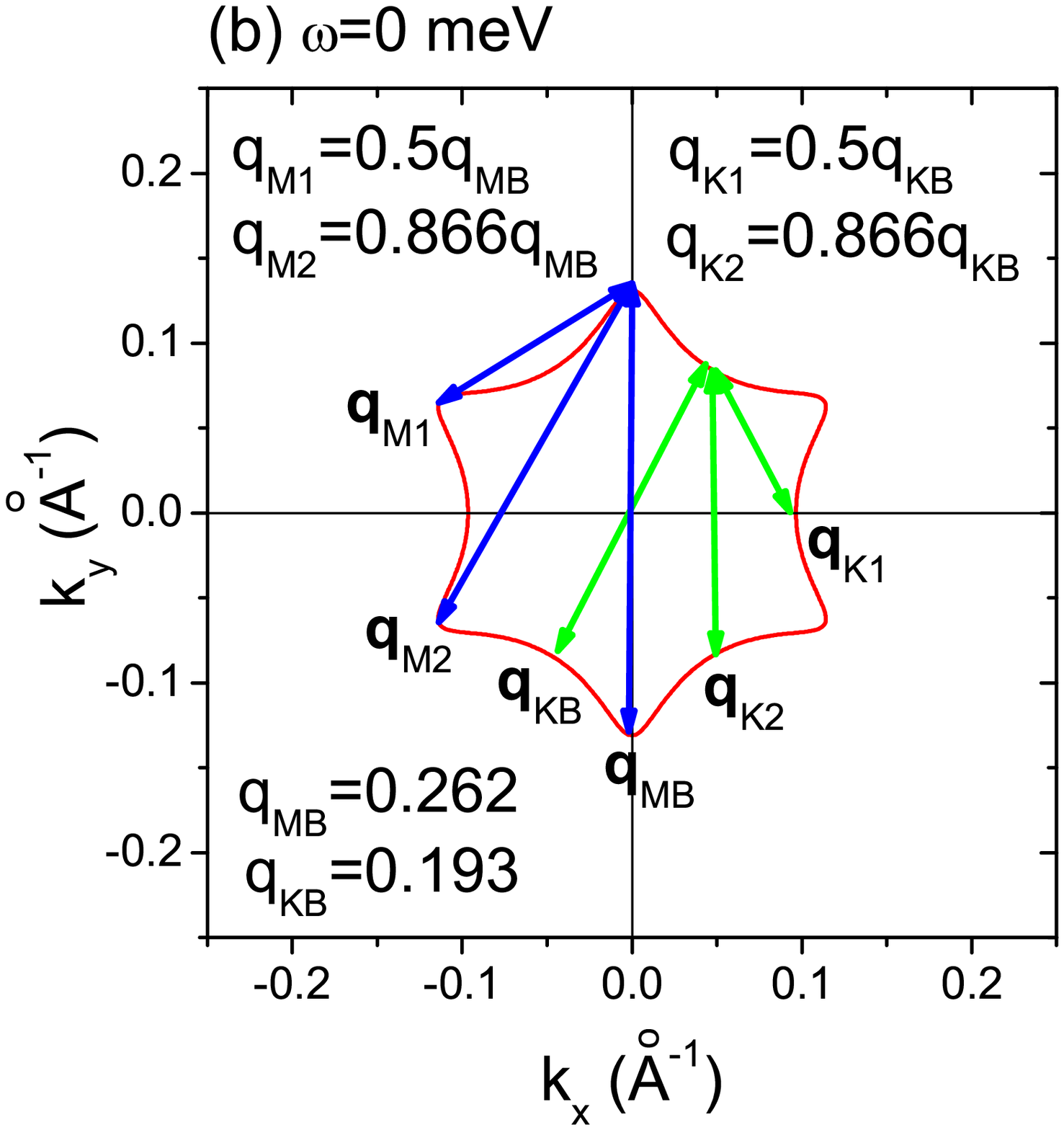}}
\rotatebox[origin=c]{0}{\includegraphics[angle=0,
           height=2.2in]{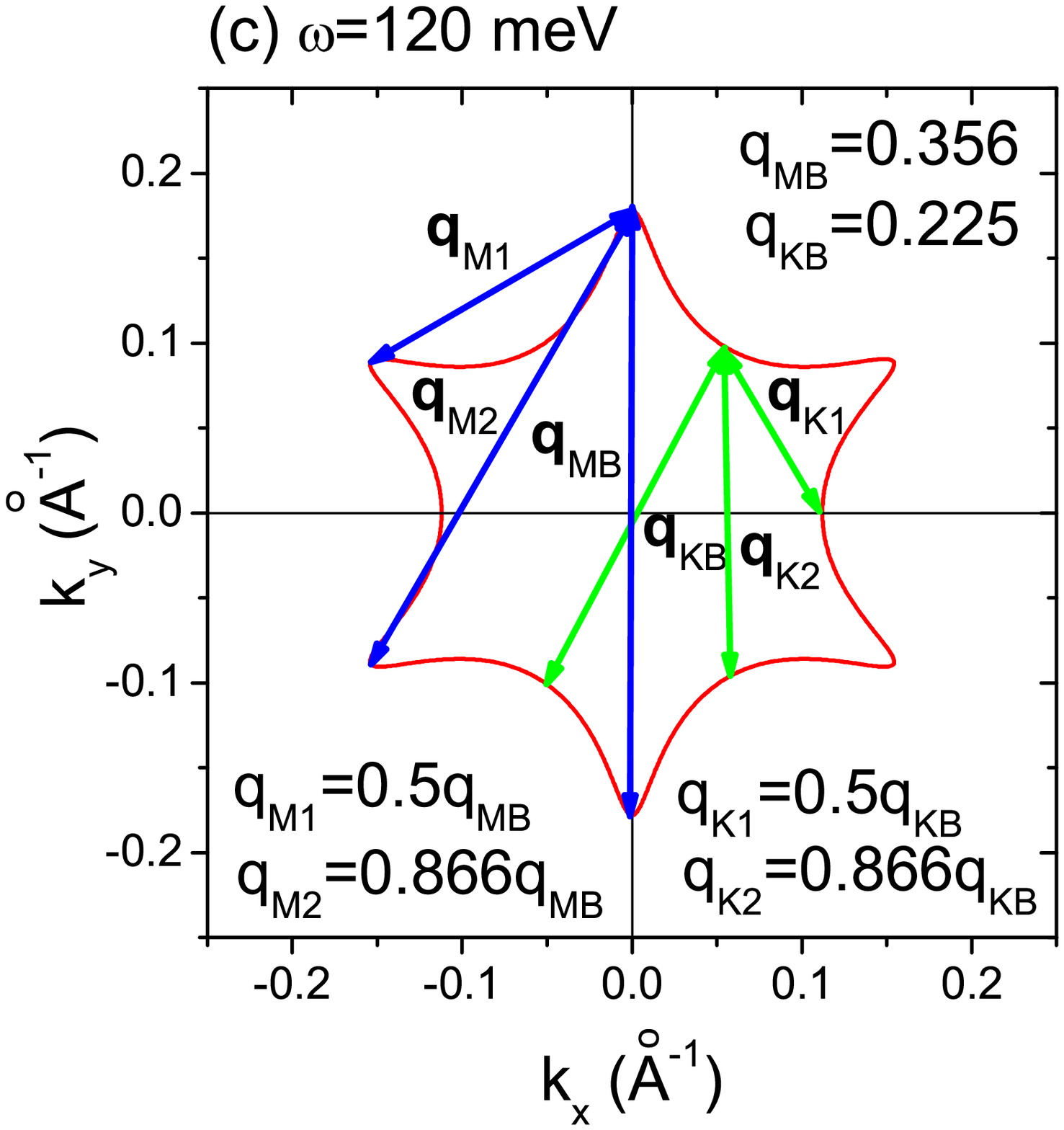}}
\caption{(Color online) The constant-energy contours of the surface
state band at different energies. The modulation wave vectors in the LDOS
are shown.}
\end{figure}

In order to understand clearly the energy-dependent peaks and dips
in the LDOS, we depict the constant-energy contours of the surface energy band at
$\omega=-120, 0, 120$ (meV) in Fig. 2. There are six modulation wave vectors
${\bf q}_{K1}$, ${\bf q}_{K2}$, ${\bf q}_{KB}$,
${\bf q}_{M1}$, ${\bf q}_{M2}$, ${\bf q}_{MB}$, and their corresponding
symmetric wave vectors induced by the QPI.
${\bf q}_{KB}$ and ${\bf q}_{MB}$ are the backscattering wave vectors connecting
two opposite K points and two opposite M points, respectively.
According to the symmetry of the constant-energy contours,
we have $|{\bf q}_{K1}|=\frac{1}{2}|{\bf q}_{KB}|$,
$|{\bf q}_{K2}|=\frac{\sqrt{3}}{2}|{\bf q}_{KB}|$,
$|{\bf q}_{M1}|=\frac{1}{2}|{\bf q}_{MB}|$, and
$|{\bf q}_{M2}|=\frac{\sqrt{3}}{2}|{\bf q}_{MB}|$.
We also note that the wave vectors ${\bf q}_{K1}$, ${\bf q}_{KB}$, and ${\bf q}_{M2}$
(${\bf q}_{M1}$, ${\bf q}_{MB}$, and ${\bf q}_{K2}$) are along the same modulation
direction, i.e. ${\bf q}_{K1}//{\bf q}_{KB}//{\bf q}_{M2}$
(${\bf q}_{M1}//{\bf q}_{MB}//{\bf q}_{K2}$).
The angle between the two modulation directions is $integer\cdot
60^\circ+30^\circ$.
Due to the existence of a backscattering wave vector
and two QPI wave vectors in the same direction,
we shall observe below that the peaks and dips in the images of the LDOS
seem not to locate exactly at these modulation wave vectors.
When $\omega=-120$ meV, $|{\bf q}_{M2}|\approx |{\bf q}_{KB}|$.
With increasing the energy, all the modulation wave vectors become longer,
but the increasing length of each wave vector is different.
Therefore, the competition among the wave vectors produces abundant energy-dependent
structures in the LDOS. The oscillation periods corresponding to these
wave vectors have the relations: $T_{K1}=\frac{2\pi}{|{\bf q}_{K1}|}
=2T_{KB}$, $T_{K2}=\frac{2\pi}{|{\bf q}_{K2}|}=\frac{2}{\sqrt{3}}T_{KB}$,
$T_{M1}=\frac{2\pi}{|{\bf q}_{M1}|}
=2T_{MB}$, and $T_{M2}=\frac{2\pi}{|{\bf q}_{M2}|}=\frac{2}{\sqrt{3}}T_{MB}$.

\begin{figure}

\rotatebox[origin=c]{0}{\includegraphics[angle=0,
           height=2.2in]{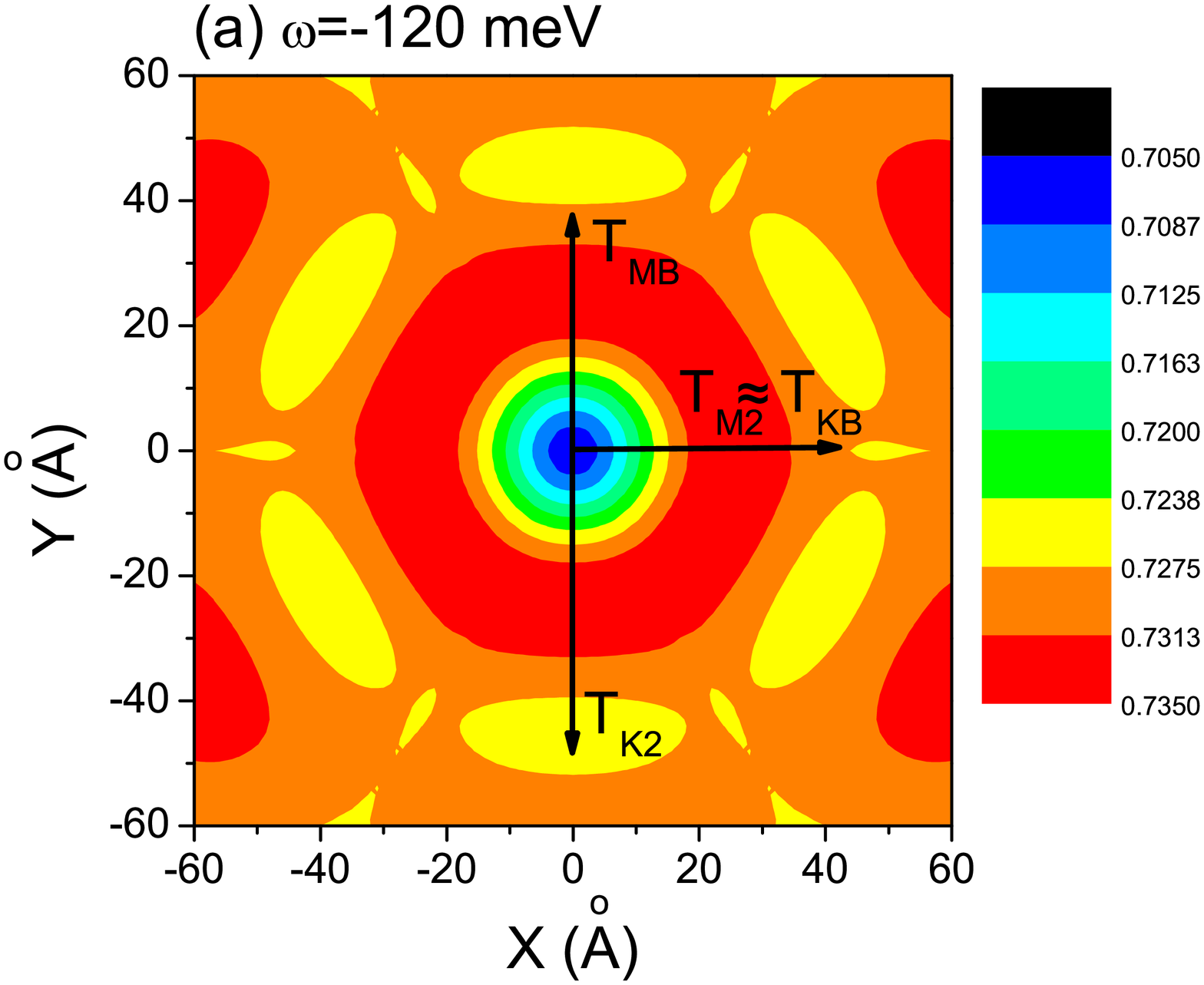}}
\rotatebox[origin=c]{0}{\includegraphics[angle=0,
           height=2.2in]{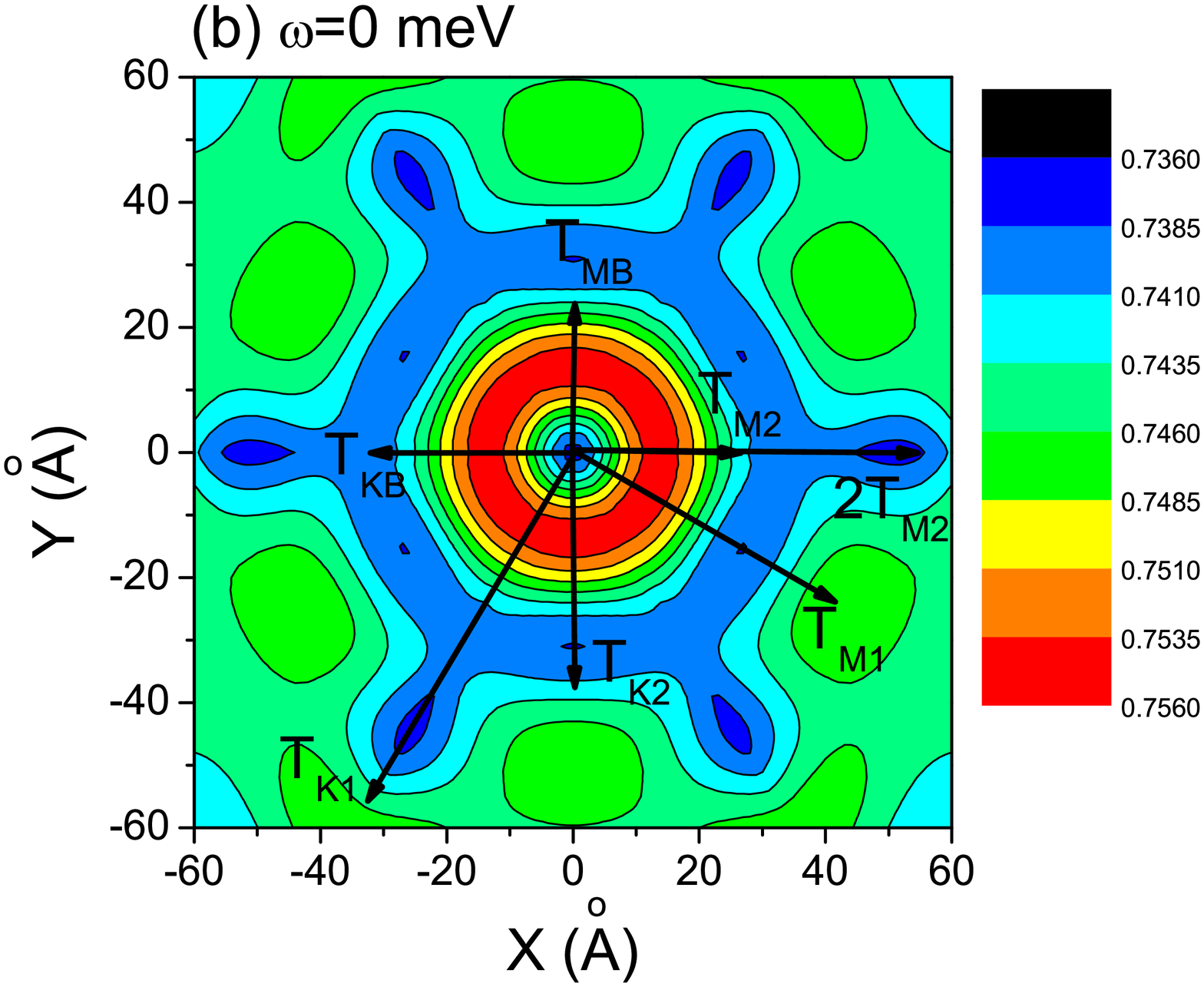}}
\rotatebox[origin=c]{0}{\includegraphics[angle=0,
           height=2.2in]{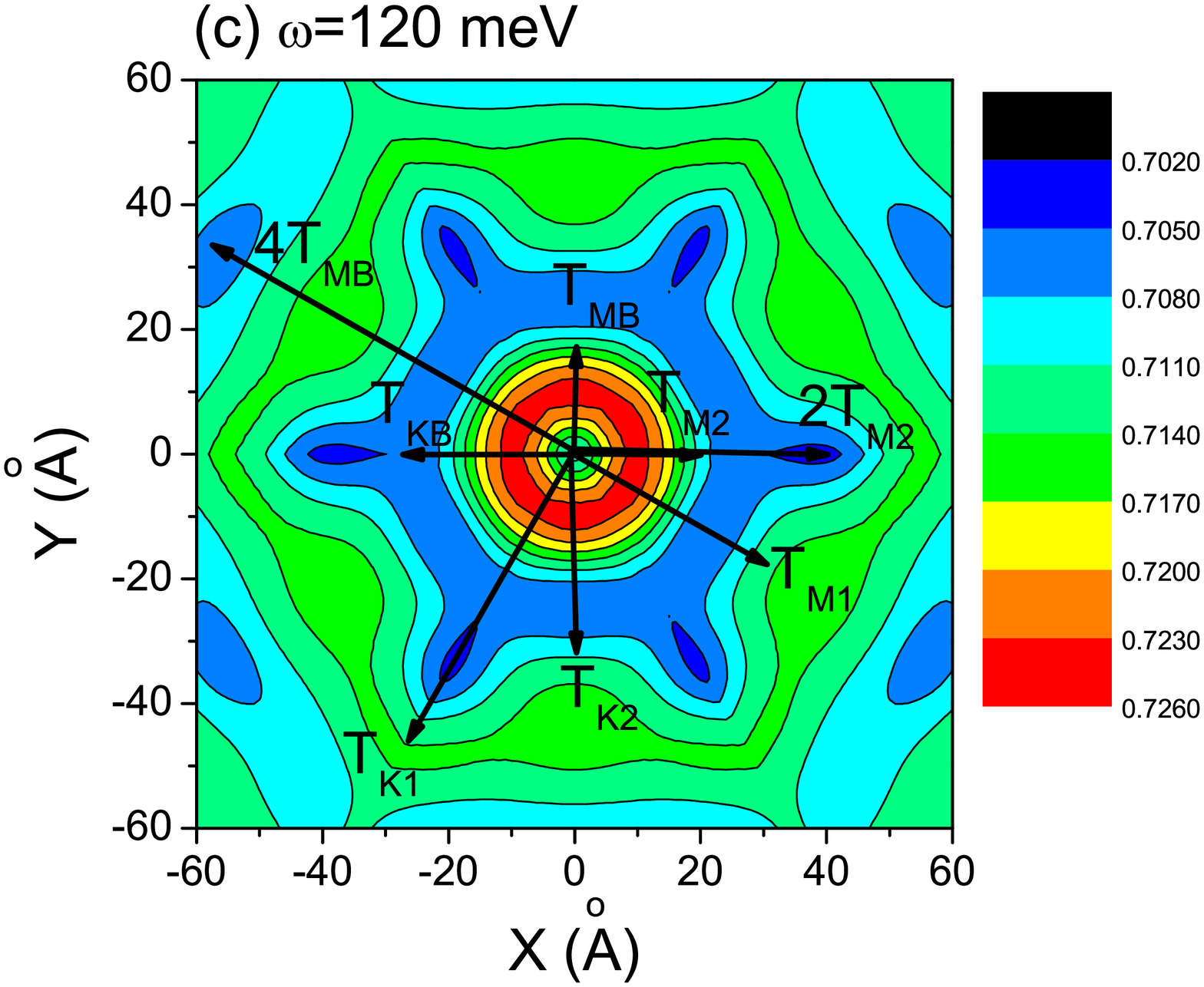}}
\caption{(Color online) The images of the LDOS with the pure nonmagnetic potential
$U=5.0$ eV at different bias voltages.}
\end{figure}

Fig. 3 shows the images of the LDOS with a moderate nonmagnetic potential $U=5.0$ eV
and $V_m=0$ eV at $\omega=-120, 0, 120$ (meV), where the TRS
persists. When $\omega=-120$ meV, the LDOS $\rho({\bf r},\omega)$ has the local minima
at $r=T_{K2}, T_{M2}, T_{KB}$, and $T_{MB}$. Because $|T_{M2}|\approx |T_{KB}|$
and $|T_{K2}|\sim |T_{MB}|$, these dips are produced by both the QPI
and the absence of backscatterings. With increasing the bias voltage,
the local minima move forward to the origin and form a hexagonal sinkage
around the impurity at $\omega=0$ meV. Obviously,
$\rho({\bf r},\omega)$ also has the peaks at $r=T_{K1}$ and $T_{M1}$ while
the dips show up at $2T_{M2}$ in Fig. 3(b). Because $T_{K1}=2T_{KB}$ and $T_{M1}=2T_{MB}$,
the effect of the QPI on the LDOS is stronger than that of the
TRS near the impurity. However, when $\omega=120$ meV, the LDOS possesses
the local minima at $r=4T_{MB}=2T_{M1}$, shown in Fig. 3(c).
This means that the effect of the TRS on the LDOS is dominant far from the impurity.
Therefore, the Friedel oscillations induced by the QPI decay faster
than those due to the TRS.

\begin{figure}

\rotatebox[origin=c]{0}{\includegraphics[angle=0,
           height=2.2in]{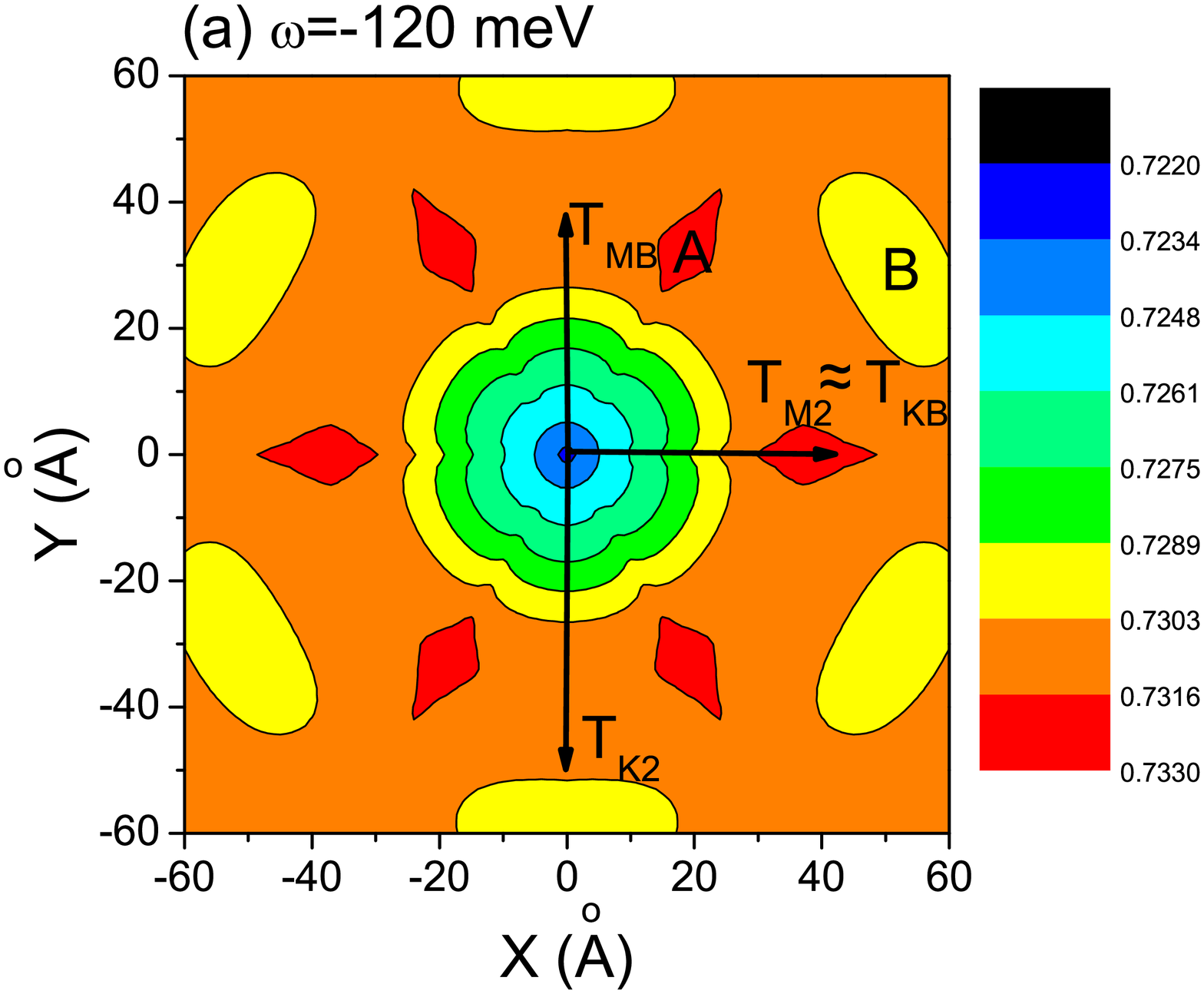}}
\rotatebox[origin=c]{0}{\includegraphics[angle=0,
           height=2.2in]{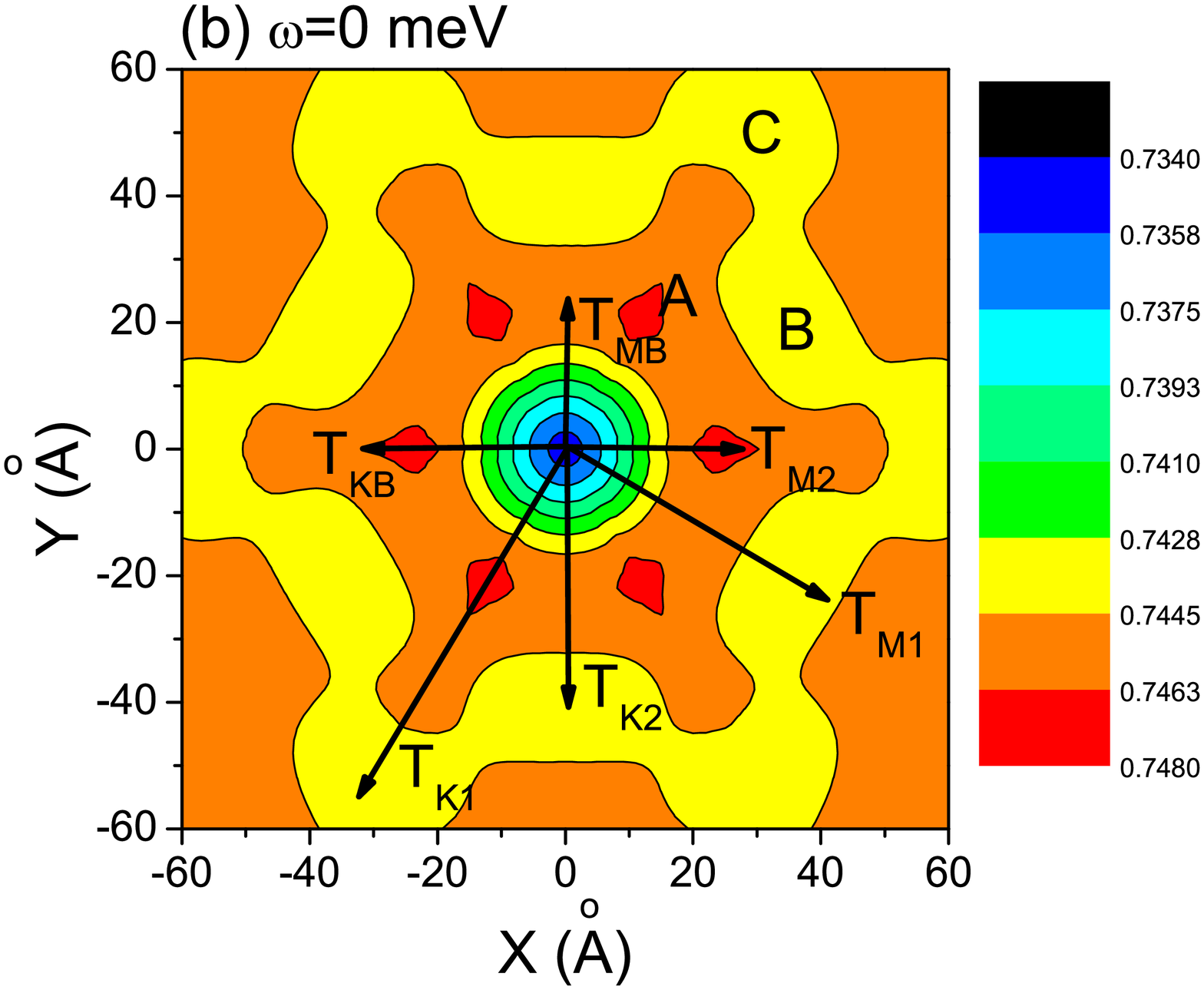}}
\rotatebox[origin=c]{0}{\includegraphics[angle=0,
           height=2.2in]{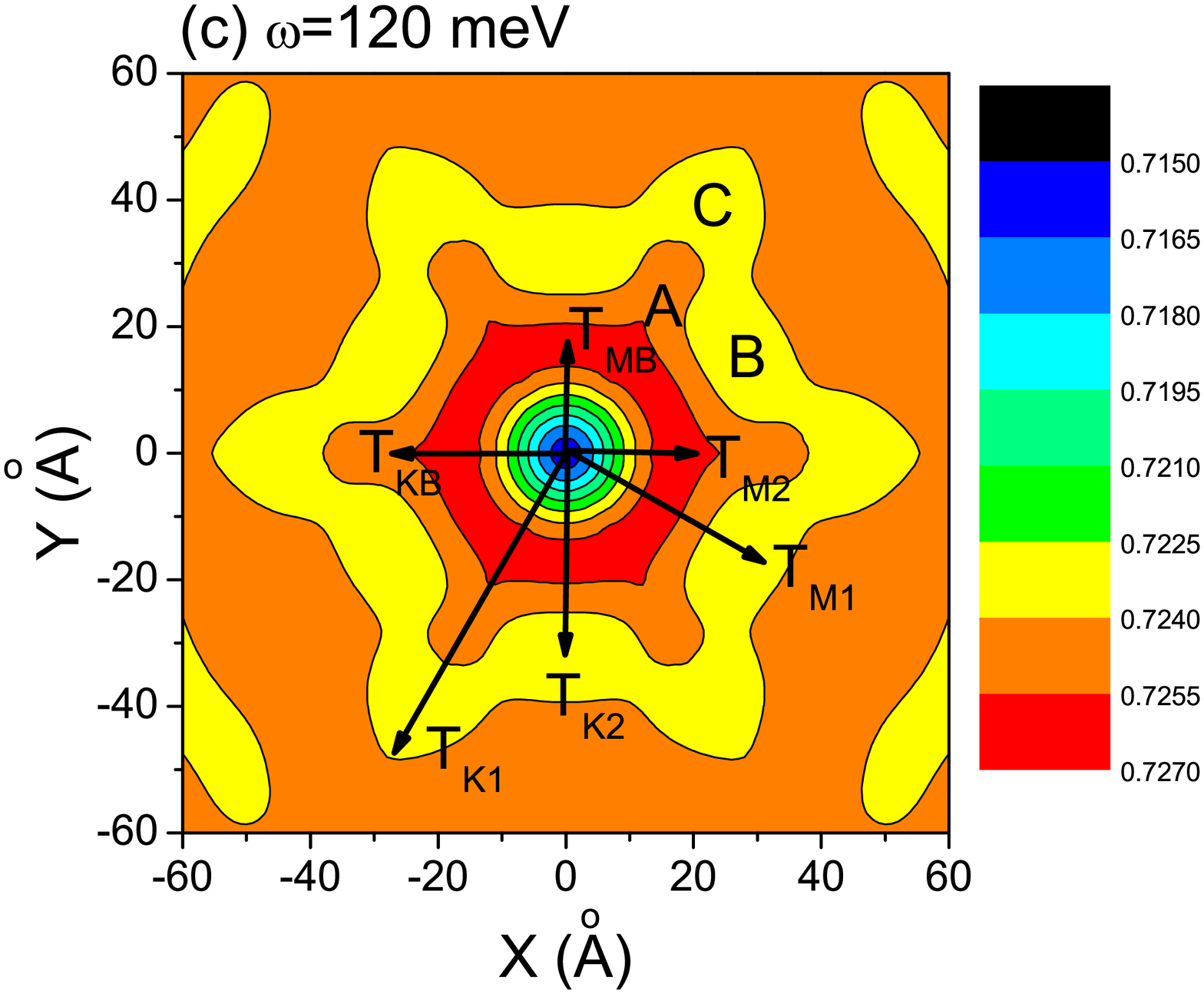}}
\caption{(Color online) The images of the LDOS with the pure magnetic potential
$V_m=10.0$ eV at different bias voltages.}
\end{figure}

To further elucidate the origin of the dips at the backscattering
wave vectors, we also calculate the LDOS produced by a magnetic
potential $V_m=10.0$ eV and $U=0$ eV, where the TRS
is broken. Therefore, the backscatterings are allowed.
When $\omega=-120$ meV, the LDOS has higher peaks at $r=T_{M2}$, opposite
with a pure nonmagnetic impurity, while there are also the dips at
$r=T_{K2}$. However, $\rho({\bf r},\omega)$ has no obvious peaks or dips
at ${\bf q}_{MB}$ and ${\bf q}_{KB}$. So the QPI can be
neglected at the backscattering wave vectors. Therefore, the dips at the
backscattering wave vectors are indeed produced by the TRS
in the presence of a nonmagnetic impurity. When the energy increases,
the peaks and the dips also move forward to the magnetic impurity,
similar to a nonmagnetic potential. In Fig. 4(b) and 4(c),
the LDOS has the local minima at $r=T_{K1}$ and $T_{M1}$, contrary to
a nonmagnetic impurity.

In summary, we have proposed a method to detect the TRS
in topological surface states. Due to this symmetry, the topological surface
states with opposite momentum and the same $s$ are incoherent and
contribute zero to the LDOS.
Therefore, the dips at the backscattering wave vectors
in the LDOS induced by a nonmagnetic impurity can be regarded a signature
of the TRS. We note that in each modulation direction there is the competition
between a backscattering wave vector and two QPI
wave vectors, which leads to robust dip-hump features in the LDOS.
Because the peaks induced by the QPI
decay faster than the dips produced by the TRS,
it is easy to observe the local minima at the backscattering wave
vectors at higher bias voltages and several periods away from the
nonmagnetic impurity by STM experiments.

This work was supported by the Sichuan Normal University,
by the Texas Center for Superconductivity at
the University of Houston, and by the Robert A. Welch Foundation
under the Grant no. E-1411.

\end{document}